\PassOptionsToPackage{table}{xcolor}
\documentclass[11pt,letterpaper]{article}
\usepackage[colorinlistoftodos]{todonotes} % add todo notes in the text
\usepackage{geometry, diagbox,tabularx}
\usepackage{amsmath}
\usepackage{amsthm}
\usepackage{comment}

\usepackage{amsfonts}
\newtheorem{definition}{Definition}
\newtheorem{assumption}{Assumption}
\newtheorem{theorem}{Theorem}
\newtheorem{lemma}{Lemma}
\newtheorem{remark}{Remark}
\usepackage[colorinlistoftodos]{todonotes} % add todo notes in the text
\usepackage{booktabs}
\usepackage{hyperref}
\usepackage{natbib}
\newcommand{\B}{\mathcal{B}}
\newcommand{\D}{\mathcal{D}}
\usepackage{amsmath,amssymb,booktabs}
\usepackage[table]{xcolor}
\usepackage{tikz}
\usetikzlibrary{arrows.meta}

\definecolor{budgetblue}{RGB}{38,93,145}
\definecolor{budgetred}{RGB}{166,63,62}
\title{Random Utility Models in the Consumption-Budget Setting:
Vertex and Network-Flow Characterizations\footnote{This paper was written with the assistance of ChatGPT and Claude Code, which were used to suggest mathematical arguments, assist with drafting and revising the manuscript, and help implement the code. }}
\author{
Haoge Chang\thanks{Department of Economics, Columbia University, \href{mailto:hc3516@columbia.edu}{\nolinkurl{hc3516@columbia.edu}}}
\and
Kota Saito\thanks{Division of Humanities and Social Science, California Institute of
Technology, \href{mailto:saito@caltech.edu}{\nolinkurl{saito@caltech.edu}}}
\and
Mingxi Zhu\thanks{Scheller College of Business, Georgia Institute of Technology, \href{mailto:mingxi.zhu@scheller.gatech.edu}{\nolinkurl{mingxi.zhu@scheller.gatech.edu
}}}
}

\begin{document}
\maketitle
%\todo[inline]{Arya: 1. write programs that verify Theorem 1 in a more complicated setup; 2. check that the linear optimization using the Smuedler's IP indeed equivalent (in value) to the BF on the multi-arc graph.
%}
\section{Introduction}
This note supplements our paper \textit{Alternative Computational Formulations for Testing Discrete-Choice Random Utility Models: Half-Space and Network-Flow Formulations} and serves two purposes:
\begin{enumerate}
    \item We clarify the conceptual distinctions between two classes of random utility models (RUM): those in the consumption-budget setting \citep{kitamura2019nonparametric} (hereafter KS) and those in the finite-alternative setting (e.g., Chapter~7 of \cite{chambers2016revealed}).

In particular, we show (Theorem~\ref{thm:projection}) that
each rational type, as defined in KS, can be obtained as a
coordinate projection of the deterministic choice vector
induced by a strict preference ranking that satisfies what
we call a \textbf{\textit{patch-monotonicity}} condition.

    \item We generalize the network-flow characterization for finite-alternative RUM in \cite{fiorini2004short} to the consumption-budget setting (Theorem \ref{thm:network}). Specifically, we show that a vector of choice probabilities in the consumption-budget setting is rationalizable in the KS sense if and only if it is the image of a unit flow on a \textbf{\textit{multi-arc powerset digraph}} under a fixed linear map. The key distinction is that our digraph allows multiple arcs between the same pair of nodes, whereas the digraph in Fiorini has at most one arc between any pair of nodes.

    This characterization allows us to adapt the Flow-CG algorithm developed in our paper to the consumption-budget setting.\footnote{Part of Section 4 is incorporated into our main paper.}
\end{enumerate}

The remainder of this note is organized into four sections. Section~\ref{section:KS} reviews the KS setup. For the finite-alternative RUM setup, we refer readers to our main paper. Section~\ref{section:v} presents the projection results, and Section~\ref{section:f} develops the network-flow characterization. Section \ref{section:proofs} includes all the proofs. Throughout the note, we prioritize detailed explanations and illustrative examples over brevity.

\section{Review of the Setup in KS}\label{section:KS}
\subsection{Budgets and Patches}
The material in this section is adapted from Section~3 of KS.
The primitive data in KS consist of choice frequencies observed on $J$ linear budgets, each of the form
\begin{equation}
    \B_j = \{y \in \mathbb{R}^K_{+}: p_j^\top y = 1\},
\end{equation}
where $K$ denotes the number of commodities and $p_j$ is the normalized price vector.\footnote{The original price vector is divided by the corresponding expenditure level, thereby normalizing expenditure to one.} It is assumed that the price vectors $\{p_j\}_{j=1}^J$ are strictly positive and distinct so that they induce different budget sets. 

KS partition the union of the budget sets into \textit{patches}, defined as follows.
\begin{definition}[Definition~3.2 in KS]
Let $\mathcal{X}=\{x_1,\ldots,x_I\}$ be the coarsest partition of $\bigcup_{j=1}^J \B_j$ such that, for every $i\in\{1,\ldots,I\}$ and $j\in\{1,\ldots,J\}$, the patch $x_i$ lies entirely on, strictly above, or strictly below the budget plane $\B_j$. Equivalently, two points $y_1,y_2\in\bigcup_{j=1}^J \B_j$ belong to the same patch if and only if
\[
    \operatorname{sg}(p_j^\top y_1-1)
    =
    \operatorname{sg}(p_j^\top y_2-1)
    \quad \text{for all } j=1,\ldots,J,
\]
where $\operatorname{sg}$ denotes the sign function.
\end{definition}

Intuitively, each budget is divided into patches according to whether its points lie on, strictly above, or strictly below each of the other budget planes. Each budget $\B_j$ is a disjoint union of $I_j$ patches, which we enumerate as $x_{1|j},\ldots,x_{I_j|j}$. 

The data relevant for rationalizability in the KS sense can be summarized by the probabilities of selecting each patch on each budget.

\begin{definition}[Choice probability vector; Definition~3.4 in KS]\label{def:cp_system}
The vector representation of a choice probability system is
\begin{equation}
    \rho :=
    \left(
        \rho_{1|1},\ldots,\rho_{I_1|1},
        \ldots,
        \rho_{1|J},\ldots,\rho_{I_J|J}
    \right)^\top,
\end{equation}
where $\rho_{i|j}$ is the probability that the bundle chosen from budget $\B_j$ belongs to patch $x_{i|j}$. 
\end{definition}

\begin{figure}[htbp]
    \centering
    \begin{tikzpicture}[
        scale=1.2,
        >=stealth,
        patch/.style={line width=1.6pt, line cap=round}
    ]
        % Commodity axes.
        \draw[->] (0,0) -- (4.6,0) node[below] {$y_1$};
        \draw[->] (0,0) -- (0,4.6) node[left] {$y_2$};

        % Budget intercepts and their common intersection.
        \coordinate (A1) at (0,2);
        \coordinate (B1) at (4,0);
        \coordinate (A2) at (0,4);
        \coordinate (B2) at (2,0);
        \coordinate (C) at ({4/3},{4/3});

        % Segment patches on budget 1.
        \draw[patch, blue!75!black]
            (A1) -- node[midway, above, yshift=6pt]
            {$x_{1|1}$} (C);
        \draw[patch, teal!80!black]
            (C) -- node[midway, above right]
            {$x_{2|1}$} (B1);

        % Segment patches on budget 2.
        \draw[patch, purple!75!black]
            (A2) -- node[midway, above right]
            {$x_{1|2}$} (C);
        \draw[patch, orange!85!black]
            (C) -- node[midway, left]
            {$x_{2|2}$} (B2);

        % Budget labels.
        \node[above right] at (3.6,0.2) {$\B_1$};
        \node[right] at (0.2,3.6) {$\B_2$};

        % Shared intersection patch.
        \draw[thin, black!65] (C) -- (2.05,1.95)
            node[above right, text=black]
            {$x_{3|1}=x_{3|2}$};
        \fill[white] (C) circle[radius=3.5pt];
        \fill[black] (C) circle[radius=2.3pt];
    \end{tikzpicture}
    \caption{The five distinct patches induced by two budgets with two
    commodities. The four colored segments exclude the intersection,
    which forms the shared singleton patch $x_{3|1}=x_{3|2}$.}
    \label{fig:budgets}
\end{figure}
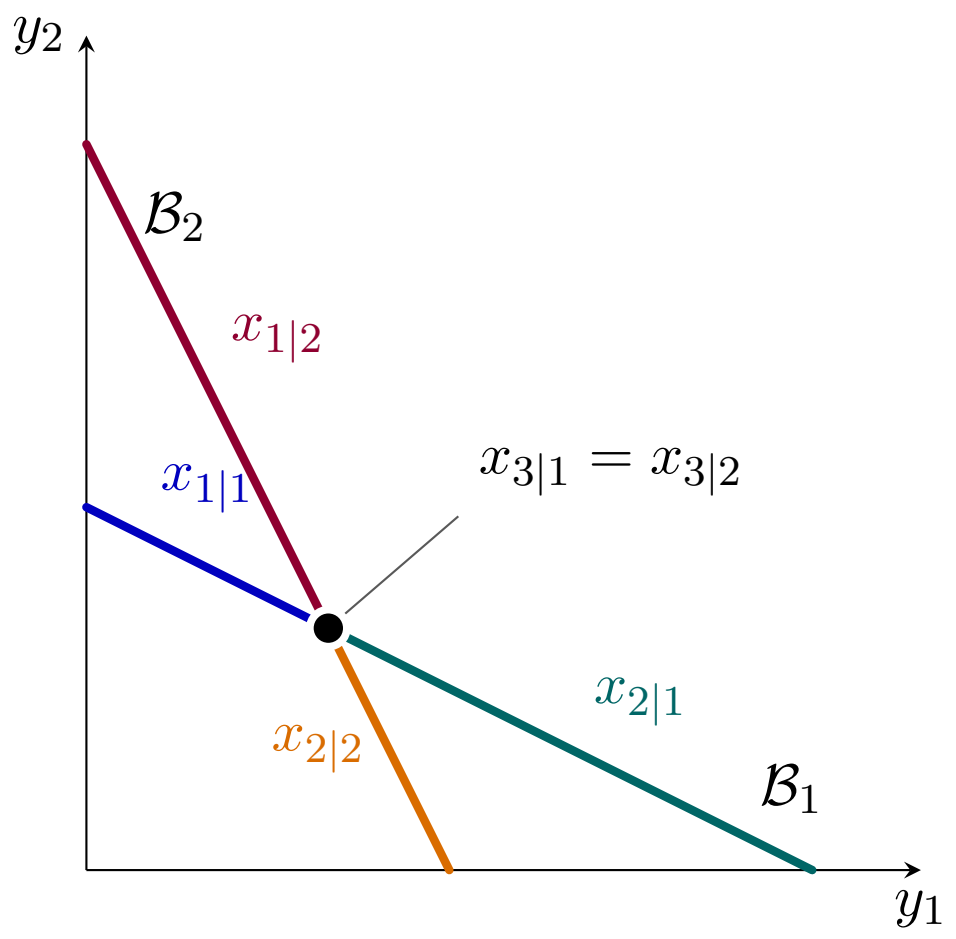

Figure \ref{fig:budgets} below provides an example of the patch constructions for a 2-commodity-2-budget case.  
\subsection{Rational Types}
The KS approach tests whether the observed vector of choice probabilities $\pi$ can be expressed as a convex combination of the deterministic choice vectors induced by \textit{rational types}. To define a rational type, we first define a \textit{choice type}. A choice type selects one patch from each budget. We represent choice type $t$ by the binary vector
\begin{equation}\label{def:type}
    a_t = \left(
        d_{1|1},\ldots,d_{I_1|1},\,
        d_{1|2},\ldots,d_{I_2|2},\,
        \ldots,\,
        d_{1|J},\ldots,d_{I_J|J}
    \right)^\top
    \in \{0,1\}^{\sum_{j=1}^J I_j},
\end{equation}
where $d_{i|j}=1$ if type $t$ selects patch $x_{i|j}$ from budget $\B_j$ and $d_{i|j}=0$ otherwise. In particular,
\begin{equation*}
    \sum_{i=1}^{I_j} d_{i|j}=1
    \qquad \text{for all } j\in\{1,\ldots,J\}.
\end{equation*}

Rationality imposes additional restrictions on a choice type, which can be formulated using either SARP or GARP. Following the simplification discussed in Section~3.2 of KS, we impose the following assumption.
\begin{assumption}[Continuous demand]\label{a:continuous_demand}
For each budget $\B_j$, the distribution of demand is continuous.\footnote{Here, continuity is understood as absolute continuity with respect to the $(K-1)$-dimensional surface measure on $\B_j$. See Section~3 of \cite{kitamura2019nonparametric} for the definition of a stochastic demand system.}
\end{assumption}
With pairwise distinct budget hyperplanes, intersection patches---those contained in at least two budgets---have lower dimension than the budget hyperplanes. Under Assumption~\ref{a:continuous_demand}, these patches have zero choice probability and can be omitted from the analysis. On the remaining patches, SARP and GARP impose the same restrictions on choice types.

From now on, all definitions involving patches exclude intersection patches, and we use SARP to define rationality. We retain the notation $x_{1|j},\ldots,x_{I_j|j}$ for the remaining patches on budget $\B_j$. Following Section~3.4.1 of KS, a rationality type is defined with respect to a directed graph on budgets.

\begin{definition}[Induced Revealed Preference Relation on Budgets]\label{def:rp}
Fix a choice type $a_t$ and the collection of budgets $\B=\{\B_1,...,\B_J\}$. For any pair $j,k\in \{1,...,J\}$, we say that $\B_k$ is directly revealed preferred to $\B_j$ if 
there is an index $i\in\{1,\ldots,I_j\}$ such that
\begin{equation}
    d_{i|j}=1
    \quad\text{and}\quad
    p_k^\top y<1
    \quad\text{for all } y\in x_{i|j}.
\end{equation}
Moreover, we define a directed revealed-preference graph with node set $\B=\{\B_1,\ldots,\B_J\}$. For any distinct budgets $\B_j$ and $\B_k$, include an arc $\B_k\to\B_j$ if $\B_k$ is revealed preferred to $\B_j$.
\end{definition}

By construction, all points in a patch lie on the same side of each budget hyperplane, so the induced relation does not depend on the particular point chosen within a patch. A choice type satisfies SARP if and only if its induced graph is acyclic, yielding the following definition.

\begin{definition}[Rational Type]\label{def:rationaltype}
A choice type $a_t$ is a \textit{rational type} if its induced revealed preference graph on budgets is acyclic.
\end{definition}

Given a collection of budgets, we can construct the collection of all distinct rational types $\{a_t\}_{t=1}^T$, where $T$ is the number of such types. 
Recall that $I$ denotes the number of patches, which is equivalent to the dimension of the choice probability vector.
Let $I$ denote the dimension of the choice probability vector. We stack the rational types columnwise to form a matrix $A\in \mathbb{R}^{I\times T}$. By Theorem~3.1 of KS, a choice probability vector $\rho$ is rationalizable if and only if
\begin{equation*}
    \exists v\in \Delta^T \text{ such that }\rho = Av,
\end{equation*}
where $\Delta^T$ is the unit simplex in $\mathbb{R}^T$, $\Delta^T=\{x\in \mathbb{R}^T_+: \sum_{t=1}^T x_t=1\}$. 

We denote the polytope generated by the KS rational types by
$\mathcal{P}^{\mathrm{KS}}$:
\begin{equation}\label{def:KS}
    \mathcal{P}^{\mathrm{KS}}
    :=
    \{\rho\in\mathbb{R}^I:
      \rho=Av \text{ for some } v\in\Delta^T\}.
\end{equation}

\section{Patch Monotonicity and a Vertex Characterization}\label{section:v}

\subsection{Preliminary Discussion}
KS's approach employs discretization. In particular, KS partition the union of the budgets into finitely many mutually exclusive patches.

If we treat each patch as an individual alternative, we can consider a finite-alternative RUM that allows all $I!$ strict rankings over the patches, where $I$ is the number of patches. We expand the choice set associated with each budget to include all patches lying on or below the corresponding budget hyperplane. That is, for each $\B_j$, we define the choice set
\begin{equation}\label{eqn:choice_set}
    \D_j=\left\{x\in\mathcal{X}: x\subseteq \{y\in\mathbb{R}^K_{+}: p_j^\top y\leq 1\}\right\}. 
\end{equation}
Here, $\mathcal{X}$ is the set of patches (alternatives) 
 The $J$ budgets index a choice probability system $\{\{\rho(x,\D_j)\}_{x\in\D_j}\}_{j=1}^J$, with one choice probability vector for each budget $\B_j$.

A choice probability vector $\{\rho(x,\D_j)\}_{x\in\D_j}$ induces a subprobability vector on the budget boundary $\B_j$, given by $\{\rho(x,\D_j)\}_{x\in\D_j:\,x\subseteq\B_j}$. This need not be a valid choice probability vector in the KS setting because its entries may sum to less than one. More generally, some of the $I!$ strict rankings do not induce rational types in the consumption-budget setting.\footnote{The need to exclude some strict preference rankings also arises when each choice set consists only of patches on the corresponding budget boundary.} Consider again the example with two budgets and two commodities.

For the budgets $\B_1,\B_2$ in Figure~\ref{fig:budgets}, the expanded menus (excluding intersection patches) are
\[
    \D_1=\{x_{1|1},x_{2|1},x_{2|2}\},\qquad
    \D_2=\{x_{1|2},x_{2|2},x_{1|1}\}.
\]
There are three rational types in the sense of KS, as shown in Table~\ref{tab:ks_types}.

\begin{table}[htbp]
    \centering
    \begin{tabular}{lccc}
        \toprule
        Budget & Type 1 & Type 2 & Type 3\\
        \midrule
        $\B_1$ & $x_{1|1}$ & $x_{2|1}$ & $x_{2|1}$\\
        $\B_2$ & $x_{1|2}$ & $x_{1|2}$ & $x_{2|2}$\\
        \bottomrule
    \end{tabular}
    \caption{The three rational types.}
    \label{tab:ks_types}
\end{table}

The remaining boundary-choice profile $(x_{1|1},x_{2|2})$ violates SARP:
$x_{2|2}$ lies strictly below $\B_1$, and $x_{1|1}$ lies
strictly below $\B_2$, so each chosen patch is directly revealed
preferred to the other. 

Table~\ref{tab:rum_types} lists all 24 strict rankings over the four patches and their induced
choices from the expanded menus. Eight of these rankings (colored in gray) induce rational
types in the KS setting: two generate Type~1, four generate Type~2, and
two generate Type~3. The remaining rankings select a patch strictly below
the corresponding budget hyperplane for at least one budget. The column denoted by monotonicity will be used in the next subsection.

\begin{table}[!htb]
    \centering
    \small
    \setlength{\tabcolsep}{5pt}
    \begin{tabular}{@{}lccc@{}}
        \toprule
        Ranking & Choice from $\D_1$ & Choice from $\D_2$ & Monotone?\\
        \midrule
        $x_{1|1}\succ x_{2|1}\succ x_{1|2}\succ x_{2|2}$ & $x_{1|1}$ & $x_{1|1}$ & No\\
        $x_{1|1}\succ x_{2|1}\succ x_{2|2}\succ x_{1|2}$ & $x_{1|1}$ & $x_{1|1}$ & No\\
        $x_{1|1}\succ x_{1|2}\succ x_{2|1}\succ x_{2|2}$ & $x_{1|1}$ & $x_{1|1}$ & No\\
        $x_{1|1}\succ x_{1|2}\succ x_{2|2}\succ x_{2|1}$ & $x_{1|1}$ & $x_{1|1}$ & No\\
        $x_{1|1}\succ x_{2|2}\succ x_{2|1}\succ x_{1|2}$ & $x_{1|1}$ & $x_{1|1}$ & No\\
        $x_{1|1}\succ x_{2|2}\succ x_{1|2}\succ x_{2|1}$ & $x_{1|1}$ & $x_{1|1}$ & No\\
        $x_{2|1}\succ x_{1|1}\succ x_{1|2}\succ x_{2|2}$ & $x_{2|1}$ & $x_{1|1}$ & No\\
        $x_{2|1}\succ x_{1|1}\succ x_{2|2}\succ x_{1|2}$ & $x_{2|1}$ & $x_{1|1}$ & No\\
        \rowcolor{gray!25}
        $x_{2|1}\succ x_{1|2}\succ x_{1|1}\succ x_{2|2}$ & $x_{2|1}$ & $x_{1|2}$ & Yes\\
        \rowcolor{gray!25}
        $x_{2|1}\succ x_{1|2}\succ x_{2|2}\succ x_{1|1}$ & $x_{2|1}$ & $x_{1|2}$ & Yes\\
        \rowcolor{gray!25}
        $x_{2|1}\succ x_{2|2}\succ x_{1|1}\succ x_{1|2}$ & $x_{2|1}$ & $x_{2|2}$ & No\\
        \rowcolor{gray!25}
        $x_{2|1}\succ x_{2|2}\succ x_{1|2}\succ x_{1|1}$ & $x_{2|1}$ & $x_{2|2}$ & Yes\\
        \rowcolor{gray!25}
        $x_{1|2}\succ x_{1|1}\succ x_{2|1}\succ x_{2|2}$ & $x_{1|1}$ & $x_{1|2}$ & Yes\\
        \rowcolor{gray!25}
        $x_{1|2}\succ x_{1|1}\succ x_{2|2}\succ x_{2|1}$ & $x_{1|1}$ & $x_{1|2}$ & No\\
        \rowcolor{gray!25}
        $x_{1|2}\succ x_{2|1}\succ x_{1|1}\succ x_{2|2}$ & $x_{2|1}$ & $x_{1|2}$ & Yes\\
        \rowcolor{gray!25}
        $x_{1|2}\succ x_{2|1}\succ x_{2|2}\succ x_{1|1}$ & $x_{2|1}$ & $x_{1|2}$ & Yes\\
        $x_{1|2}\succ x_{2|2}\succ x_{1|1}\succ x_{2|1}$ & $x_{2|2}$ & $x_{1|2}$ & No\\
        $x_{1|2}\succ x_{2|2}\succ x_{2|1}\succ x_{1|1}$ & $x_{2|2}$ & $x_{1|2}$ & No\\
        $x_{2|2}\succ x_{1|1}\succ x_{2|1}\succ x_{1|2}$ & $x_{2|2}$ & $x_{2|2}$ & No\\
        $x_{2|2}\succ x_{1|1}\succ x_{1|2}\succ x_{2|1}$ & $x_{2|2}$ & $x_{2|2}$ & No\\
        $x_{2|2}\succ x_{2|1}\succ x_{1|1}\succ x_{1|2}$ & $x_{2|2}$ & $x_{2|2}$ & No\\
        $x_{2|2}\succ x_{2|1}\succ x_{1|2}\succ x_{1|1}$ & $x_{2|2}$ & $x_{2|2}$ & No\\
        $x_{2|2}\succ x_{1|2}\succ x_{1|1}\succ x_{2|1}$ & $x_{2|2}$ & $x_{2|2}$ & No\\
        $x_{2|2}\succ x_{1|2}\succ x_{2|1}\succ x_{1|1}$ & $x_{2|2}$ & $x_{2|2}$ & No\\
        \bottomrule
    \end{tabular}
    \caption{All strict rankings and their induced choices from the expanded menus.
    Shaded rows induce rational types in the KS setting.
    The final column indicates whether the ranking satisfies patch monotonicity.}
    \label{tab:rum_types}
\end{table}

A question is therefore how we can restrict the strict preference rankings such that they induce a KS-stype rational types on the boundary patches. It turns out that one sufficient condition exists and we term it \textbf{\textit{patch monotonicity}}.

\subsection{Patch Monotonicity and the Projection Theorem}

\begin{definition}[Patch Monotonicity]\label{def:patch_monotonicity}
For distinct patches $x_1,x_2\in\mathcal{X}$, we write $x_1\succ^d x_2$ if there exist $y_1\in x_1$ and $y_2\in x_2$ such that $y_1\geq y_2$ componentwise.\footnote{For $y_1,y_2\in\mathbb{R}^K_+$, we say that $y_1$ dominates $y_2$ if $y_{1k}\geq y_{2k}$ for every $k\in\{1,\ldots,K\}$, with strict inequality for at least one coordinate.} We say that a strict preference ranking $\succ$ satisfies patch monotonicity if $x_1\succ^d x_2$ implies $x_1\succ x_2$ for all distinct $x_1,x_2\in\mathcal{X}$.\footnote{\cite{koida2024dual} introduce a notion of \textit{dominance} (and \textit{undominance}). The two notions differ: theirs is a relation between a patch and a budget, whereas $\succ^d$ is a relation between two patches. They are nonetheless connected: if $x'\succ^d x$ for some patch $x'$ on budget $\B_k$, then $x$ lies strictly below $\B_k$, and hence is dominated by $\B_k$ in the sense of \cite{koida2024dual}.}

\end{definition}

\begin{remark}
Table~\ref{tab:rum_types} illustrates that satisfying patch monotonicity is sufficient, but not necessary, for a strict ranking to induce a rational type in the sense of KS. Nevertheless, each rational type in this example is induced by at least one strict ranking that satisfies patch monotonicity.
\end{remark}

\begin{remark}
One may ask why Definition~\ref{def:patch_monotonicity} uses an existential rather than a universal quantifier (i.e., every point in one patch is dominated by some point
in the other patch). Consider the example with three budgets and two commodities in Figure~\ref{fig:three_budget_dominance}. The budget $\B_1$ lies strictly inside the feasible budget sets associated with $\B_2$ and $\B_3$. Each patch on these outer budgets contains a point that dominates some point of $x_{1|1}$. However, neither left patch contains a point dominating $(5/3,0)\in x_{1|1}$, and neither right patch contains a point dominating $(0,5/3)\in x_{1|1}$. Thus, no single outer patch can provide a dominating point for every point of $x_{1|1}$.

A version of patch monotonicity requiring such a universal comparison would allow the ranking
\[
    x_{1|1}\succ x_{2|2}\succ x_{1|3}\succ x_{1|2}\succ x_{2|3}.
\]
This ranking selects $x_{1|1}$ from both expanded menus $\D_2$ and $\D_3$ and therefore does not induce a rational type in the sense of KS.

\begin{figure}[htbp]
    \centering
    \begin{tikzpicture}[x=1.33cm,y=1.33cm,>=Stealth,
        every node/.style={font=\small}]

        % Inner feasible budget set
        \fill[gray!8]
            (0,0)--({5/3},0)--(0,{5/3})--cycle;

        % Axes
        \draw[->,gray!85] (0,0)--(4.65,0)
            node[right] {$y_1$};
        \draw[->,gray!85] (0,0)--(0,4.55)
            node[above] {$y_2$};

        \foreach \t in {1,2,4} {
            \draw (\t,0.045)--(\t,-0.045)
                node[below=3pt] {$\t$};
            \draw (0.045,\t)--(-0.045,\t)
                node[left=3pt] {$\t$};
        }
        \node[below left=3pt] at (0,0) {$0$};

        % Budget boundaries
        \draw[very thick]
            (0,{5/3})--({5/3},0);
        \draw[very thick,budgetblue]
            (0,2)--(4,0);
        \draw[very thick,budgetred]
            (0,4)--(2,0);

        % Patch labels
        \node[fill=white,inner sep=1.5pt]
            at (0.65,0.82) {$x_{1|1}$};

        \node[budgetblue,fill=white,inner sep=1.5pt]
            at (0.54,1.95) {$x_{1|2}$};
        \node[budgetblue,fill=white,inner sep=1.5pt]
            at (2.86,0.79) {$x_{2|2}$};

        \node[budgetred,fill=white,inner sep=1.5pt]
            at (0.79,2.79) {$x_{1|3}$};
        \node[budgetred,fill=white,inner sep=1.5pt]
            at (1.62,0.55) {$x_{2|3}$};

        % Budget equations: expenditure normalized to one
        \node[anchor=west,fill=white,inner sep=1.5pt]
            at (2.35,3.55)
            {$\B_1:\ \frac35 y_1+\frac35 y_2=1$};

        \node[anchor=west,budgetblue,fill=white,inner sep=1.5pt]
            at (2.35,3.15)
            {$\B_2:\ \frac14 y_1+\frac12 y_2=1$};

        \node[anchor=west,budgetred,fill=white,inner sep=1.5pt]
            at (2.35,2.75)
            {$\B_3:\ \frac12 y_1+\frac14 y_2=1$};

    \end{tikzpicture}
    \caption{The inner budget $\B_1$ is a single patch.
    Each outer budget has two patches.
    The intersection patch is excluded.}
    \label{fig:three_budget_dominance}
\end{figure}
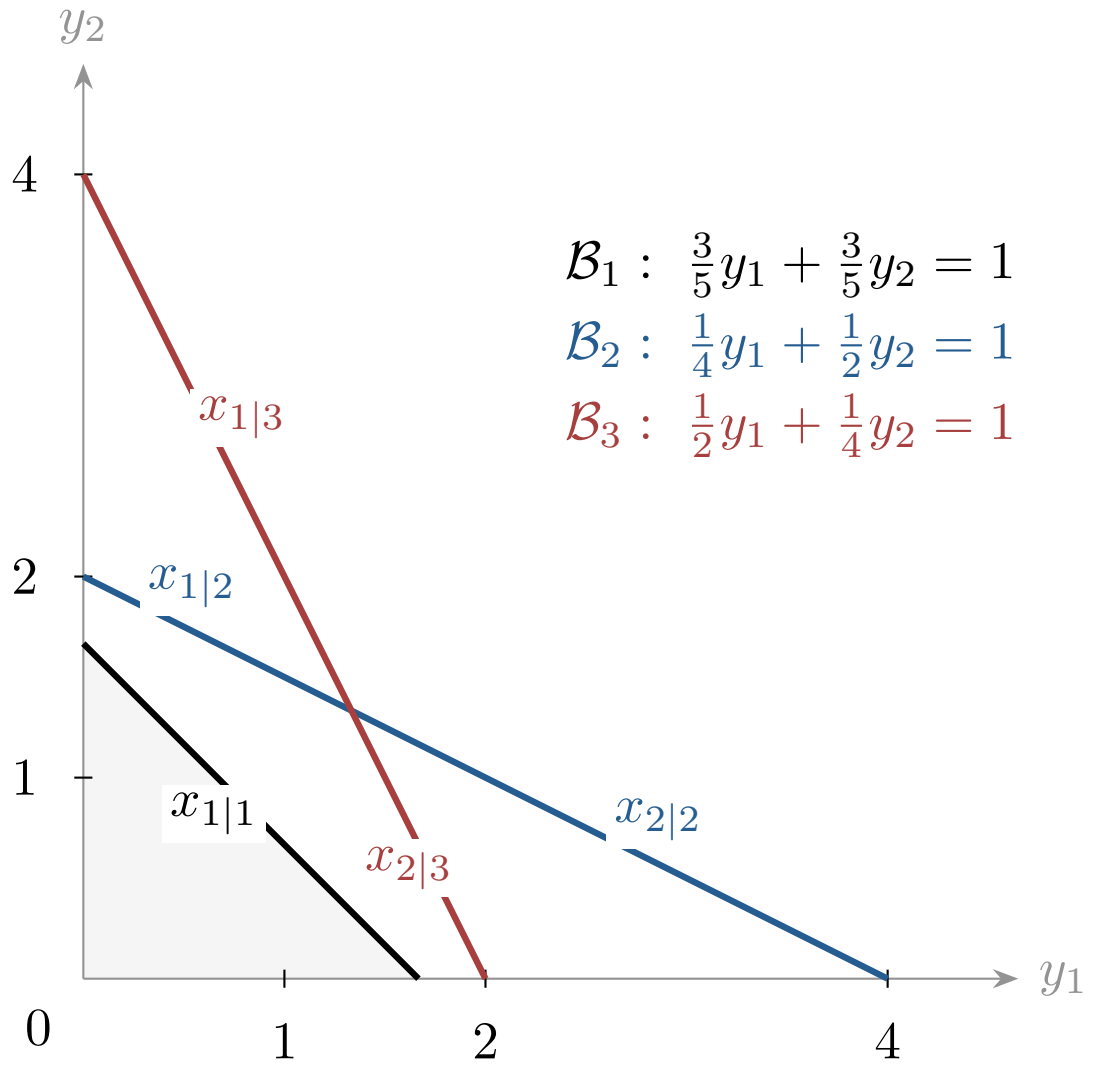
\end{remark}

We now state our projection theorem. Recall the definition of $\mathcal{P}^{\textrm{KS}}$ from \eqref{def:KS}.

\begin{theorem}\label{thm:projection}
Given a collection of distinct budgets $\{\B_j\}_{j=1}^J$, let $\mathcal{X}=\{x_1,\ldots,x_I\}$ denote the induced nonintersection patches. Let $a_1,\ldots,a_T\in\{0,1\}^I$ denote the rational choice types as in Definition \ref{def:rationaltype}, and write $A=[a_1,\ldots,a_T]$. Let $\Pi$ denote the set of all strict preference rankings over $\mathcal{X}$, and let $\Pi_r\subseteq\Pi$ denote the rankings that satisfy patch monotonicity as defined in Definition~\ref{def:patch_monotonicity}.

Let $\{\D_j\}_{j=1}^J$ be the choice sets defined in \eqref{eqn:choice_set}. For each $\succ\in\Pi$, let $\rho^\succ$ denote the deterministic choice probability vector obtained by selecting the highest-ranked patch from each $\D_j$. Then the following statements hold:
\begin{enumerate}
    \item Every ranking in $\Pi_r$ induces a rational choice type in the sense of KS.

    \item Every rational choice type in the sense of KS is induced by at least one ranking in $\Pi_r$.

    \item Define
    \[
        \mathcal{P}^r
        =
        \left\{
            \sum_{\succ\in\Pi_r}\lambda_\succ\rho^\succ:
            \lambda_\succ\geq 0
            \text{ for all }\succ\in\Pi_r,\quad
            \sum_{\succ\in\Pi_r}\lambda_\succ=1
        \right\}.
    \]
    Let $\mathrm{Proj}$ denote the coordinate projection that retains only the coordinates $\rho(x,\D_j)$ for which $x\subseteq\B_j$, ordered as in the rows of $A$. Then
    \begin{equation}
       \mathcal{P}^{\mathrm{KS}}
        =
        \mathrm{Proj}(\mathcal{P}^r),
    \end{equation}

    In words, the set of rationalizable choice probability systems in the sense of KS is the projection of the choice probability polytope generated by rankings that satisfy patch monotonicity.
\end{enumerate}
\end{theorem}

\section{Multi-arc Network Flow Characterization}\label{section:f}

In the finite-alternative RUM setting, \cite{fiorini2004short}  establishes a correspondence between strict preference rankings over $n$ alternatives and directed paths from the empty set to the full set of alternatives in the corresponding powerset digraph.

This section shows that an analogous characterization holds in the consumption-budget setting, with some modifications. In particular, we construct a powerset digraph whose nodes correspond to subsets of budgets. Since the individual choice alternatives are patches, a given ordering of the budgets may be consistent with multiple patch choices. To represent these possibilities, we use a \textbf{multi-arc powerset digraph}, in which multiple arcs between the same pair of nodes represent distinct admissible patch choices.

\subsection{Examples}

In this section, we demonstrate how to write multi-arc powerset digraphs with several examples.

\subsubsection{Two Commodities and Three Budgets}
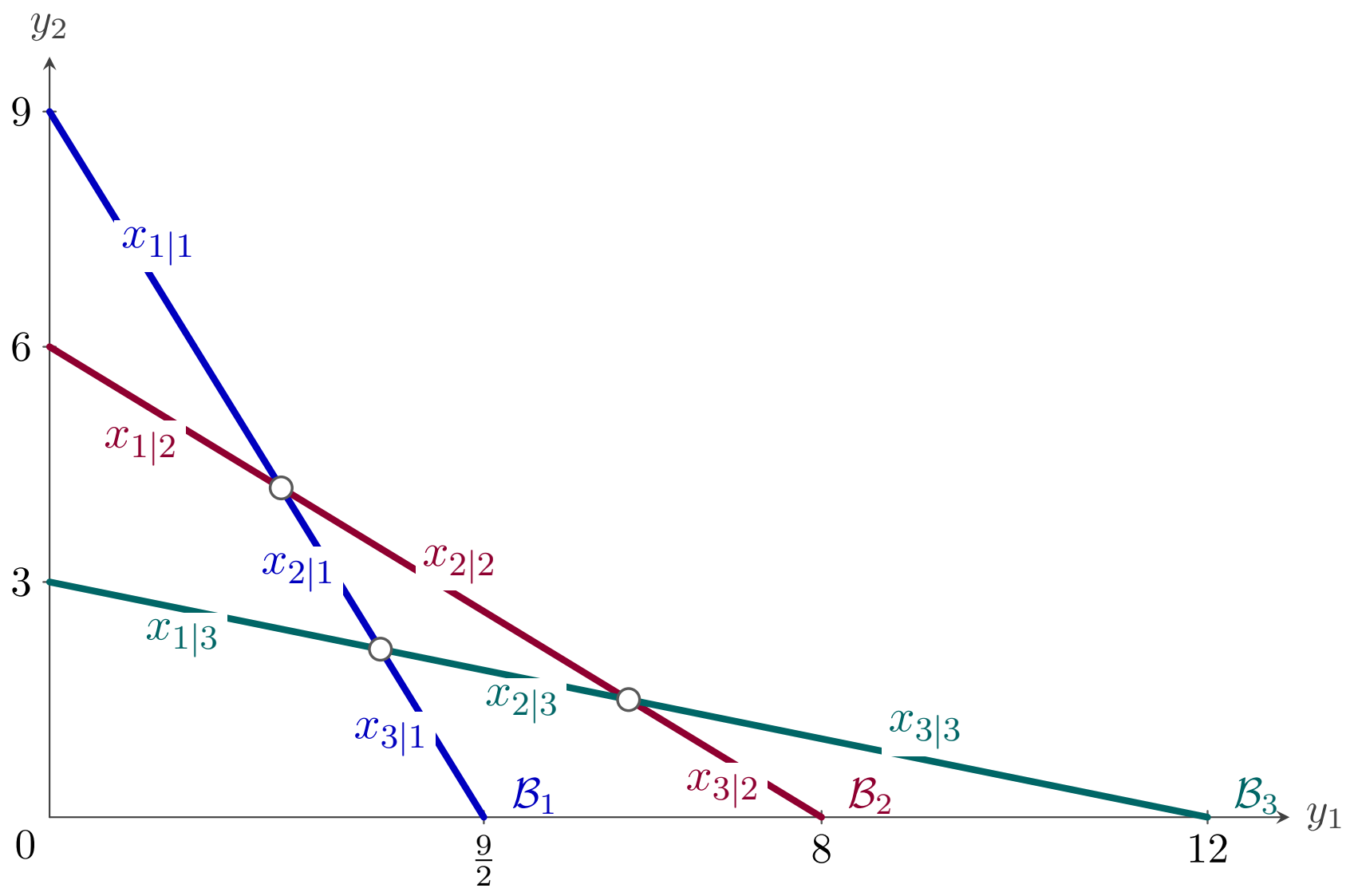
\begin{figure}[htbp]
    \centering
    \begin{tikzpicture}[
        x=0.8cm,y=0.65cm,
        >=stealth,
        line cap=round,
        line join=round,
        font=\small,
        budget/.style={line width=1.6pt},
        patchlabel/.style={
            fill=white,
            inner sep=1.6pt,
            font=\normalsize
        },
        tick/.style={line width=0.45pt,draw=black!65}
    ]
        % Commodity axes.
        \draw[->,black!75] (0,0) -- (12.85,0)
            node[right] {$y_1$};
        \draw[->,black!75] (0,0) -- (0,9.70)
            node[above] {$y_2$};
        \node[below left,inner sep=3pt] at (0,0) {$0$};

        % Budget intercepts.
        \foreach \xx/\lab in {4.5/{\frac{9}{2}},8/{8},12/{12}} {
            \draw[tick] (\xx,0.08) -- (\xx,-0.08);
            \node[below,inner sep=4pt] at (\xx,0) {$\lab$};
        }
        \foreach \yy in {3,6,9} {
            \draw[tick] (-0.065,\yy) -- (0.065,\yy);
            \node[left,inner sep=4pt] at (0,\yy) {$\yy$};
        }

        % Budget frontiers.
        \draw[budget,blue!75!black] (0,9) -- (4.5,0);
        \draw[budget,purple!75!black] (0,6) -- (8,0);
        \draw[budget,teal!80!black] (0,3) -- (12,0);

        % Excluded intersections.
        \coordinate (P12) at ({12/5},{21/5});
        \coordinate (P13) at ({24/7},{15/7});
        \coordinate (P23) at (6,{3/2});
        \foreach \P in {P12,P13,P23} {
            \filldraw[
                fill=white,
                draw=black!65,
                line width=0.65pt
            ] (\P) circle[radius=2.6pt];
        }

        % Patches on budget 1.
        \node[patchlabel,text=blue!75!black]
            at (1.13,7.28) {$x_{1|1}$};
        \node[patchlabel,text=blue!75!black]
            at (2.58,3.12) {$x_{2|1}$};
        \node[patchlabel,text=blue!75!black]
            at (3.54,1.02) {$x_{3|1}$};

        % Patches on budget 2.
        \node[patchlabel,text=purple!75!black]
            at (0.95,4.73) {$x_{1|2}$};
        \node[patchlabel,text=purple!75!black]
            at (4.25,3.22) {$x_{2|2}$};
        \node[patchlabel,text=purple!75!black]
            at (6.98,0.36) {$x_{3|2}$};

        % Patches on budget 3.
        \node[patchlabel,text=teal!80!black]
            at (1.38,2.28) {$x_{1|3}$};
        \node[patchlabel,text=teal!80!black]
            at (4.90,1.44) {$x_{2|3}$};
        \node[patchlabel,text=teal!80!black]
            at (9.08,1.11) {$x_{3|3}$};

        % Budget labels.
        \node[text=blue!75!black,anchor=west]
            at (4.62,0.27) {$\B_1$};
        \node[text=purple!75!black,anchor=west]
            at (8.10,0.27) {$\B_2$};
        \node[text=teal!80!black,anchor=west]
            at (12.10,0.27) {$\B_3$};
    \end{tikzpicture}
    \caption{The nine nonintersection patches induced by three budgets
    with two commodities. Patches on each budget are indexed in increasing
    order of $y_1$. Hollow circles mark the excluded intersection points.}
    \label{fig:patches}
\end{figure}

We begin with an example involving two commodities and three budgets.
Consider the following pairwise intersecting budgets in $\mathbb{R}_+^2$:
\begin{equation}\label{eq:budgets}
\begin{aligned}
    \B_1&=\left\{y\in\mathbb{R}_+^2:
        \frac{2}{9}y_1+\frac{1}{9}y_2=1\right\},\\
    \B_2&=\left\{y\in\mathbb{R}_+^2:
        \frac{1}{8}y_1+\frac{1}{6}y_2=1\right\},\\
    \B_3&=\left\{y\in\mathbb{R}_+^2:
        \frac{1}{12}y_1+\frac{1}{3}y_2=1\right\}.
\end{aligned}
\end{equation}
After excluding intersection patches, each budget $\B_j$ has three
patches, which we enumerate as $x_{1|j},x_{2|j},x_{3|j}$ in increasing
order of $y_1$. The pairwise intersection points are
\[
    P_{12}=\left(\frac{12}{5},\frac{21}{5}\right),\qquad
    P_{13}=\left(\frac{24}{7},\frac{15}{7}\right),\qquad
    P_{23}=\left(6,\frac{3}{2}\right).
\]
Figure~\ref{fig:patches} illustrates the resulting nine nonintersection patches.

\begin{figure}[htbp]
    \centering

    % Common styles for the three panels.
    \tikzset{
        sequence/.style={
            x=0.25cm,y=0.25cm,
            >=stealth,
            line cap=round,
            line join=round,
            font=\scriptsize
        },
        eligible/.style={
            line width=1.5pt
        },
        remaining/.style={
            line width=0.65pt,
            opacity=0.5
        },
        ineligible/.style={
            draw=gray!65,
            densely dashed,
            line width=0.7pt
        },
        patchlabel/.style={
            fill=white,
            inner sep=0.8pt
        },
        intersection/.style={
            fill=white,
            draw=black!60,
            line width=0.45pt
        }
    }

    % Common axes, dimensions, and intersection coordinates.
    \newcommand{\budgetsequenceaxes}{%
        \path[use as bounding box]
            (-0.7,-1.0) rectangle (13.6,10.5);
        \draw[->,black!65,line width=0.4pt]
            (0,0) -- (12.7,0) node[right] {$y_1$};
        \draw[->,black!65,line width=0.4pt]
            (0,0) -- (0,9.6) node[above] {$y_2$};
        \node[below left,inner sep=1.5pt] at (0,0) {$0$};
        \coordinate (P12) at ({12/5},{21/5});
        \coordinate (P13) at ({24/7},{15/7});
        \coordinate (P23) at (6,{3/2});
    }

    % Panel (a): all three budgets remain.
    \begin{minipage}[t]{0.32\linewidth}
        \centering
        {\small (a) Choose from $\B_1$\par}
        \smallskip
        \begin{tikzpicture}[sequence]
            \budgetsequenceaxes

            % Other remaining budgets.
            \draw[remaining,purple!75!black]
                (0,6) -- (8,0);
            \draw[remaining,teal!80!black]
                (0,3) -- (12,0);

            % Eligible patch on budget 1.
            \draw[eligible,blue!75!black]
                (0,9)
                -- node[midway,right,patchlabel,text=blue!75!black]
                {$x_{1|1}$} (P12);

            % Ineligible patches on budget 1.
            \draw[ineligible]
                (P12)
                -- node[midway,left,patchlabel,text=gray!75]
                {$x_{2|1}$} (P13);
            \draw[ineligible]
                (P13)
                -- node[midway,left,patchlabel,text=gray!75]
                {$x_{3|1}$} (4.5,0);

            % Original intersection points.
            \foreach \P in {P12,P13,P23} {
                \filldraw[intersection]
                    (\P) circle[radius=1.6pt];
            }

            \node[below,text=blue!75!black]
                at (4.5,0) {$\B_1$};
            \node[below,text=purple!75!black]
                at (8,0) {$\B_2$};
            \node[below,text=teal!80!black]
                at (12,0) {$\B_3$};
        \end{tikzpicture}
    \end{minipage}\hfill
    % Panel (b): budget 1 has been removed.
    \begin{minipage}[t]{0.32\linewidth}
        \centering
        {\small (b) Choose from $\B_2$\par}
        \smallskip
        \begin{tikzpicture}[sequence]
            \budgetsequenceaxes

            % Other remaining budget.
            \draw[remaining,teal!80!black]
                (0,3) -- (12,0);

            % Eligible patches on budget 2.
            \draw[eligible,purple!75!black]
                (0,6)
                -- node[
                    midway,above,xshift=2pt,
                    patchlabel,text=purple!75!black
                ] {$x_{1|2}$} (P12);
            \draw[eligible,purple!75!black]
                (P12)
                -- node[midway,above,patchlabel,text=purple!75!black]
                {$x_{2|2}$} (P23);

            % Ineligible patch on budget 2.
            \draw[ineligible] (P23) -- (8,0);
            \draw[gray!65,line width=0.3pt]
                (7,0.75) -- (7.6,1.8);
            \node[above,patchlabel,text=gray!75]
                at (7.6,1.8) {$x_{3|2}$};

            % Retain the original patch boundaries.
            \foreach \P in {P12,P13,P23} {
                \filldraw[intersection]
                    (\P) circle[radius=1.6pt];
            }

            \node[below,text=purple!75!black]
                at (8,0) {$\B_2$};
            \node[below,text=teal!80!black]
                at (12,0) {$\B_3$};
        \end{tikzpicture}
    \end{minipage}\hfill
    % Panel (c): only budget 3 remains.
    \begin{minipage}[t]{0.32\linewidth}
        \centering
        {\small (c) Choose from $\B_3$\par}
        \smallskip
        \begin{tikzpicture}[sequence]
            \budgetsequenceaxes

            % All three original patches are eligible.
            \draw[eligible,teal!80!black]
                (0,3)
                -- node[midway,above,patchlabel,text=teal!80!black]
                {$x_{1|3}$} (P13);
            \draw[eligible,teal!80!black]
                (P13)
                -- node[midway,above,patchlabel,text=teal!80!black]
                {$x_{2|3}$} (P23);
            \draw[eligible,teal!80!black]
                (P23)
                -- node[midway,above,patchlabel,text=teal!80!black]
                {$x_{3|3}$} (12,0);

            % Retain the original patch boundaries.
            \foreach \P in {P13,P23} {
                \filldraw[intersection]
                    (\P) circle[radius=1.6pt];
            }

            \node[below,text=teal!80!black]
                at (12,0) {$\B_3$};
        \end{tikzpicture}
    \end{minipage}

    \caption{Sequential choices for the budget ordering
    $\B_1\succ\B_2\succ\B_3$. Thick colored segments indicate eligible
    patches on the current budget, dashed gray segments indicate
    ineligible patches, and thin lines show the other remaining
    budgets. Hollow circles mark the original excluded intersection
    points.}
    \label{fig:budget_sequence}
\end{figure}
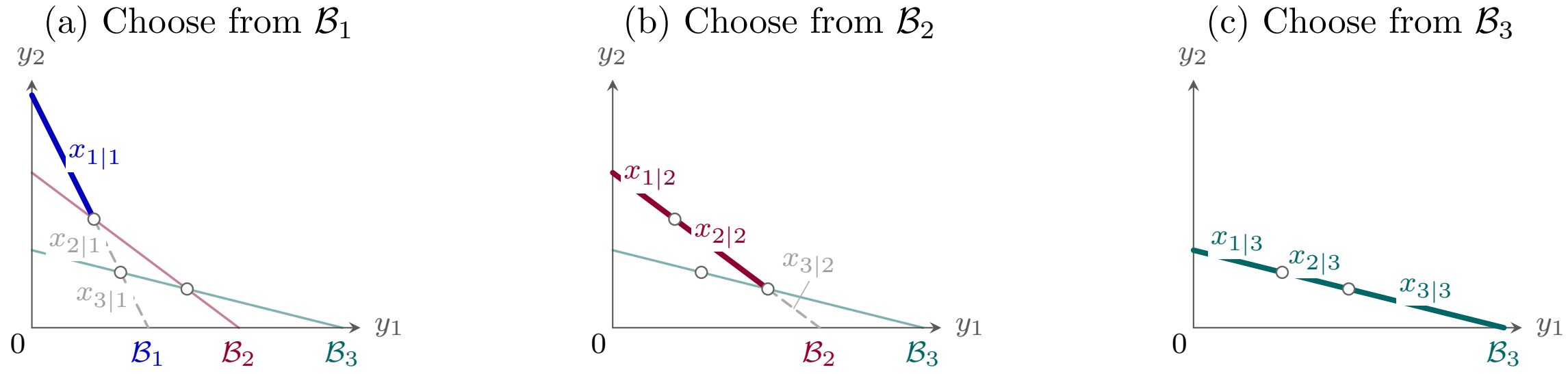

Now consider an agent whose maximized utility is highest on $\B_1$,
followed by $\B_2$ and then $\B_3$. Thus, the agent must choose $x_{1|1}$ from $\B_1$,
since $x_{2|1}$ and $x_{3|1}$ lie strictly below $\B_2$. We then remove
$\B_1$ from the construction. On $\B_2$, both $x_{1|2}$ and $x_{2|2}$
lie above the remaining budget $\B_3$, whereas $x_{3|2}$ lies strictly
below it. Either $x_{1|2}$ or $x_{2|2}$ may therefore be selected.
After removing $\B_2$, any of the three original patches on $\B_3$
may be selected. Intuitively, the patch selected from the
highest-ranked remaining budget must lie strictly above every other
remaining budget. Figure~\ref{fig:budget_sequence} illustrates this
sequence.

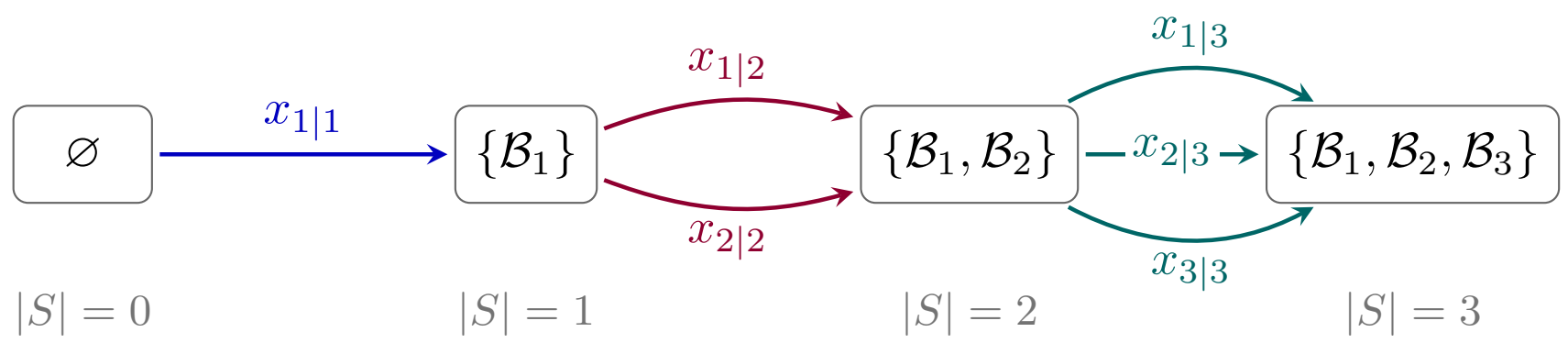
\begin{figure}[htbp]
    \centering
    \begin{tikzpicture}[
        >=stealth,
        state/.style={
            draw=black!60,fill=white,rounded corners=3pt,
            minimum width=10mm,minimum height=7mm,
            inner sep=4pt,font=\small
        },
        flow/.style={
            ->,line width=0.85pt,
            shorten >=1.4pt,shorten <=1.4pt
        },
        patchlabel/.style={
            font=\small,fill=white,inner sep=1.4pt
        },
        layerlabel/.style={
            font=\footnotesize,text=black!55
        }
    ]
        % Each node records the budgets already removed.
        \node[state] (empty) at (0,0) {$\varnothing$};
        \node[state] (one) at (3.2,0) {$\{\B_1\}$};
        \node[state] (onetwo) at (6.4,0) {$\{\B_1,\B_2\}$};
        \node[state,minimum width=16mm] (all) at (9.6,0)
            {$\{\B_1,\B_2,\B_3\}$};

        % Remove budget 1: one admissible patch.
        \draw[flow,blue!75!black] (empty) --
            node[patchlabel,midway,above=1pt]
            {$x_{1|1}$} (one);

        % Remove budget 2: two admissible patches.
        \draw[flow,purple!75!black] (one) to[bend left=18]
            node[patchlabel,midway,above=1pt]
            {$x_{1|2}$} (onetwo);
        \draw[flow,purple!75!black] (one) to[bend right=18]
            node[patchlabel,midway,below=1pt]
            {$x_{2|2}$} (onetwo);

        % Remove budget 3: three admissible patches.
        \draw[flow,teal!80!black] (onetwo) to[bend left=28]
            node[patchlabel,midway,above=1pt]
            {$x_{1|3}$} (all);
        \draw[flow,teal!80!black] (onetwo) --
            node[patchlabel,midway]
            {$x_{2|3}$} (all);
        \draw[flow,teal!80!black] (onetwo) to[bend right=28]
            node[patchlabel,midway,below=1pt]
            {$x_{3|3}$} (all);

        % Number of budgets already removed.
        \node[layerlabel] at (0,-1.15) {$|S|=0$};
        \node[layerlabel] at (3.2,-1.15) {$|S|=1$};
        \node[layerlabel] at (6.4,-1.15) {$|S|=2$};
        \node[layerlabel] at (9.6,-1.15) {$|S|=3$};
    \end{tikzpicture}
    \caption{The subnetwork corresponding to
    $\B_1\succ\B_2\succ\B_3$. Nodes record the budgets already removed.
    Arc colors identify the budget being removed, and arc labels
    identify the chosen patch.}
    \label{fig:ordered-flow}
\end{figure}

Figure~\ref{fig:ordered-flow} illustrates the subnetwork corresponding to
the budget ordering $\B_1\succ\B_2\succ\B_3$. Each node $S$ records the
indices of the budgets already removed. An arc from $S$ to
$S\cup\{\B_j\}$ represents choosing its labeled patch from $\B_j$ and then
removing $\B_j$.

The first transition has a single arc, labeled $x_{1|1}$. The second
transition has two parallel arcs, labeled $x_{1|2}$ and $x_{2|2}$,
and the final transition has three parallel arcs, one for each patch
on $\B_3$. Each path from $\varnothing$ to $\{\B_1,\B_2,\B_3\}$ selects one
patch from each budget and induces a rational type consistent with
the given ordering. The parallel arcs
capture the fact that a budget ordering can be consistent with
multiple patch choices. A deterministic rational type can thus be represented by sending one
unit of flow along a single path.

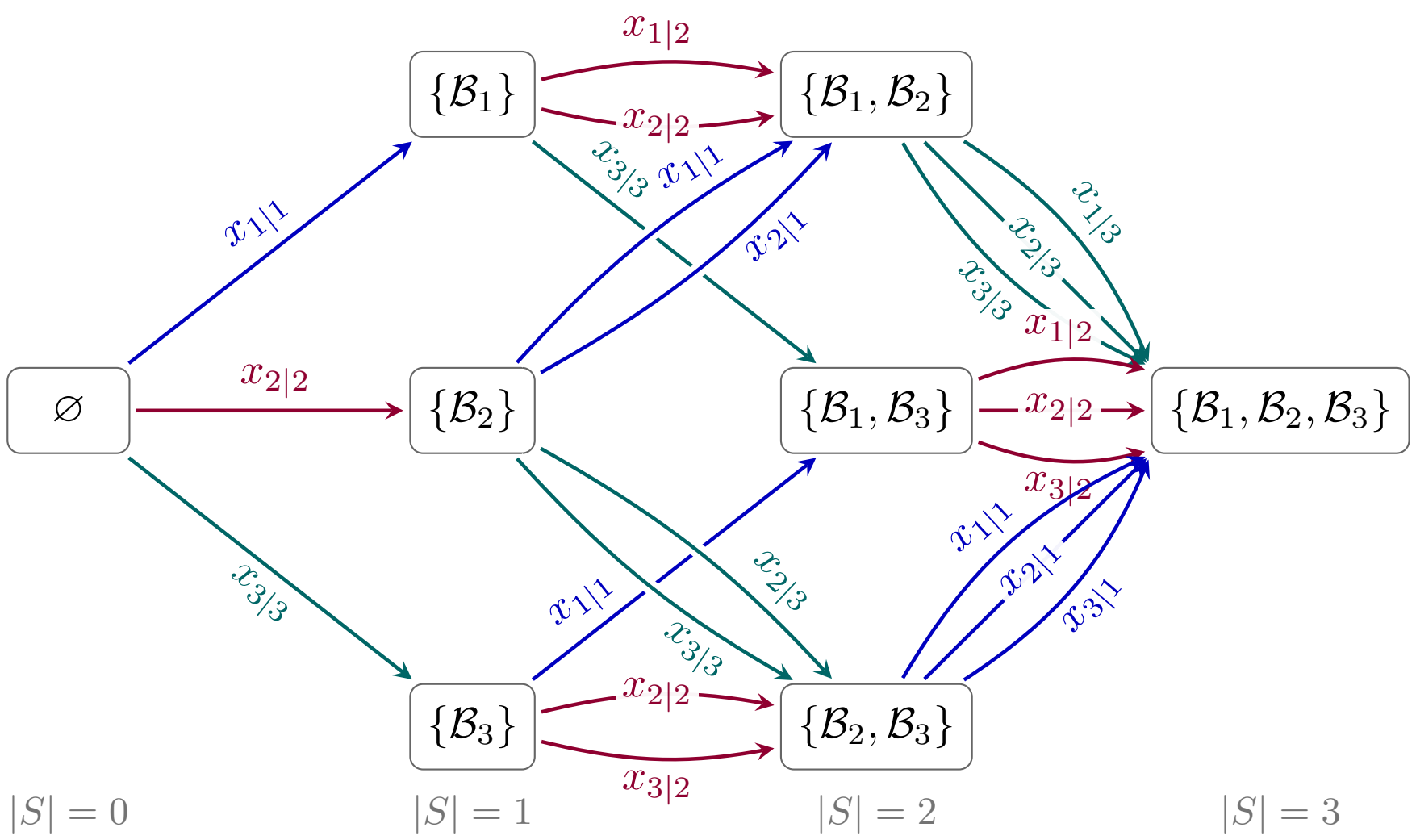
\begin{figure}[htbp]
    \centering
    \begin{tikzpicture}[
        x=0.72cm,y=0.72cm,
        >=stealth,
        state/.style={
            draw=black!60,fill=white,rounded corners=3pt,
            minimum width=10mm,minimum height=7mm,
            inner sep=4pt,font=\small
        },
        flow/.style={
            ->,line width=0.85pt,
            shorten >=1.4pt,shorten <=1.4pt
        },
        bridge/.style={
            preaction={draw=white,line width=3.5pt}
        },
        patchlabel/.style={
            font=\small,fill=white,inner sep=1.4pt,
            text opacity=1,fill opacity=0.96
        },
        layerlabel/.style={
            font=\footnotesize,text=black!55
        }
    ]
        % Nodes record the budgets already removed.
        \node[state] (empty) at (0,0) {$\varnothing$};
        \node[state] (one) at (4.6,3.6) {$\{\B_1\}$};
        \node[state] (two) at (4.6,0) {$\{\B_2\}$};
        \node[state] (three) at (4.6,-3.6) {$\{\B_3\}$};

        \node[state] (onetwo) at (9.2,3.6) {$\{\B_1,\B_2\}$};
        \node[state] (onethree) at (9.2,0) {$\{\B_1,\B_3\}$};
        \node[state] (twothree) at (9.2,-3.6) {$\{\B_2,\B_3\}$};

        \node[state,minimum width=16mm] (all) at (13.8,0)
            {$\{\B_1,\B_2,\B_3\}$};

        % Remove the first budget.
        \draw[flow,blue!75!black] (empty) --
            node[patchlabel,pos=0.52,sloped,above=1pt]
            {$x_{1|1}$} (one);
        \draw[flow,purple!75!black] (empty) --
            node[patchlabel,pos=0.52,above=1pt]
            {$x_{2|2}$} (two);
        \draw[flow,teal!80!black] (empty) --
            node[patchlabel,pos=0.52,sloped,below=1pt]
            {$x_{3|3}$} (three);

        % Draw the single diagonal arcs before the crossing arcs.
        \draw[flow,teal!80!black] (one) --
            node[patchlabel,pos=0.25,sloped,above=1pt]
            {$x_{3|3}$} (onethree);
        \draw[flow,blue!75!black] (three) --
            node[patchlabel,pos=0.25,sloped,above=1pt]
            {$x_{1|1}$} (onethree);

        % {1} -> {1,2}: two choices on budget 2.
        \draw[flow,purple!75!black] (one) to[bend left=12]
            node[patchlabel,midway,above=1pt]
            {$x_{1|2}$} (onetwo);
        \draw[flow,purple!75!black] (one) to[bend right=12]
            node[patchlabel,midway]
            {$x_{2|2}$} (onetwo);

        % {2} -> {1,2}: two choices on budget 1.
        \draw[flow,bridge,blue!75!black] (two) to[bend left=9]
            node[patchlabel,pos=0.73,sloped,above=1pt]
            {$x_{1|1}$} (onetwo);
        \draw[flow,bridge,blue!75!black] (two) to[bend right=9]
            node[patchlabel,pos=0.73,sloped,below=1pt]
            {$x_{2|1}$} (onetwo);

        % {2} -> {2,3}: two choices on budget 3.
        \draw[flow,bridge,teal!80!black] (two) to[bend left=9]
            node[patchlabel,pos=0.73,sloped,above=1pt]
            {$x_{2|3}$} (twothree);
        \draw[flow,bridge,teal!80!black] (two) to[bend right=9]
            node[patchlabel,pos=0.73,sloped,below=1pt]
            {$x_{3|3}$} (twothree);

        % {3} -> {2,3}: two choices on budget 2.
        \draw[flow,purple!75!black] (three) to[bend left=12]
            node[patchlabel,midway]
            {$x_{2|2}$} (twothree);
        \draw[flow,purple!75!black] (three) to[bend right=12]
            node[patchlabel,midway,below=1pt]
            {$x_{3|2}$} (twothree);

        % Every patch is admissible on the last remaining budget.

        % {1,2} -> {1,2,3}: three choices on budget 3.
        \draw[flow,teal!80!black] (onetwo) to[bend left=17]
            node[patchlabel,pos=0.49,sloped,above=1pt]
            {$x_{1|3}$} (all.north west);
        \draw[flow,teal!80!black] (onetwo) --
            node[patchlabel,pos=0.49,sloped]
            {$x_{2|3}$} (all.north west);
        \draw[flow,teal!80!black] (onetwo) to[bend right=17]
            node[patchlabel,pos=0.49,sloped,below=1pt]
            {$x_{3|3}$} (all.north west);

        % {1,3} -> {1,2,3}: three choices on budget 2.
        \draw[flow,purple!75!black] (onethree) to[bend left=17]
            node[patchlabel,pos=0.49,above=1pt]
            {$x_{1|2}$} (all);
        \draw[flow,purple!75!black] (onethree) --
            node[patchlabel,pos=0.49]
            {$x_{2|2}$} (all);
        \draw[flow,purple!75!black] (onethree) to[bend right=17]
            node[patchlabel,pos=0.49,below=1pt]
            {$x_{3|2}$} (all);

        % {2,3} -> {1,2,3}: three choices on budget 1.
        \draw[flow,blue!75!black] (twothree) to[bend left=17]
            node[patchlabel,pos=0.49,sloped,above=1pt]
            {$x_{1|1}$} (all.south west);
        \draw[flow,blue!75!black] (twothree) --
            node[patchlabel,pos=0.49,sloped]
            {$x_{2|1}$} (all.south west);
        \draw[flow,blue!75!black] (twothree) to[bend right=17]
            node[patchlabel,pos=0.49,sloped,below=1pt]
            {$x_{3|1}$} (all.south west);

        % Number of budgets already removed.
        \node[layerlabel] at (0,-4.6) {$|S|=0$};
        \node[layerlabel] at (4.6,-4.6) {$|S|=1$};
        \node[layerlabel] at (9.2,-4.6) {$|S|=2$};
        \node[layerlabel] at (13.8,-4.6) {$|S|=3$};
    \end{tikzpicture}
    \caption{The complete multi-arc powerset digraph for the
    three-budget example. Nodes record the budgets already removed.
    An arc's color identifies the budget being removed, and its
    label identifies the chosen patch. Crossings of arcs are not
    additional nodes.}
    \label{fig:full-flow}
\end{figure}

We complete the network by repeating this construction for the remaining
five strict rankings of the budgets. Figure~\ref{fig:full-flow} shows the complete multi-arc powerset
digraph, which has eight nodes and 22 arcs. Each path from $\varnothing$
to $\{\B_1,\B_2,\B_3\}$ specifies a budget ordering and one admissible patch choice
on each budget, thereby inducing a rational type. Different paths may
induce the same rational type.

Mapping arc flows to choice probabilities is particularly simple. For example, the probability of choosing patch $x_{1|1}$ from budget $\B_1$ equals the sum of the flows on all arcs labeled $x_{1|1}$ in Figure~\ref{fig:full-flow}.

\subsubsection{Two Commodities and Three Budgets with a Dominated Budget}

This example shows that some nodes of the powerset digraph can be
omitted, depending on the geometry of the budgets. Recall the example
in Figure~\ref{fig:three_budget_dominance}, reproduced below. Here,
every point of $\B_1$ lies strictly below the union of $\B_2$ and $\B_3$.
Thus, any feasible strict budget ordering must place $\B_1$ last

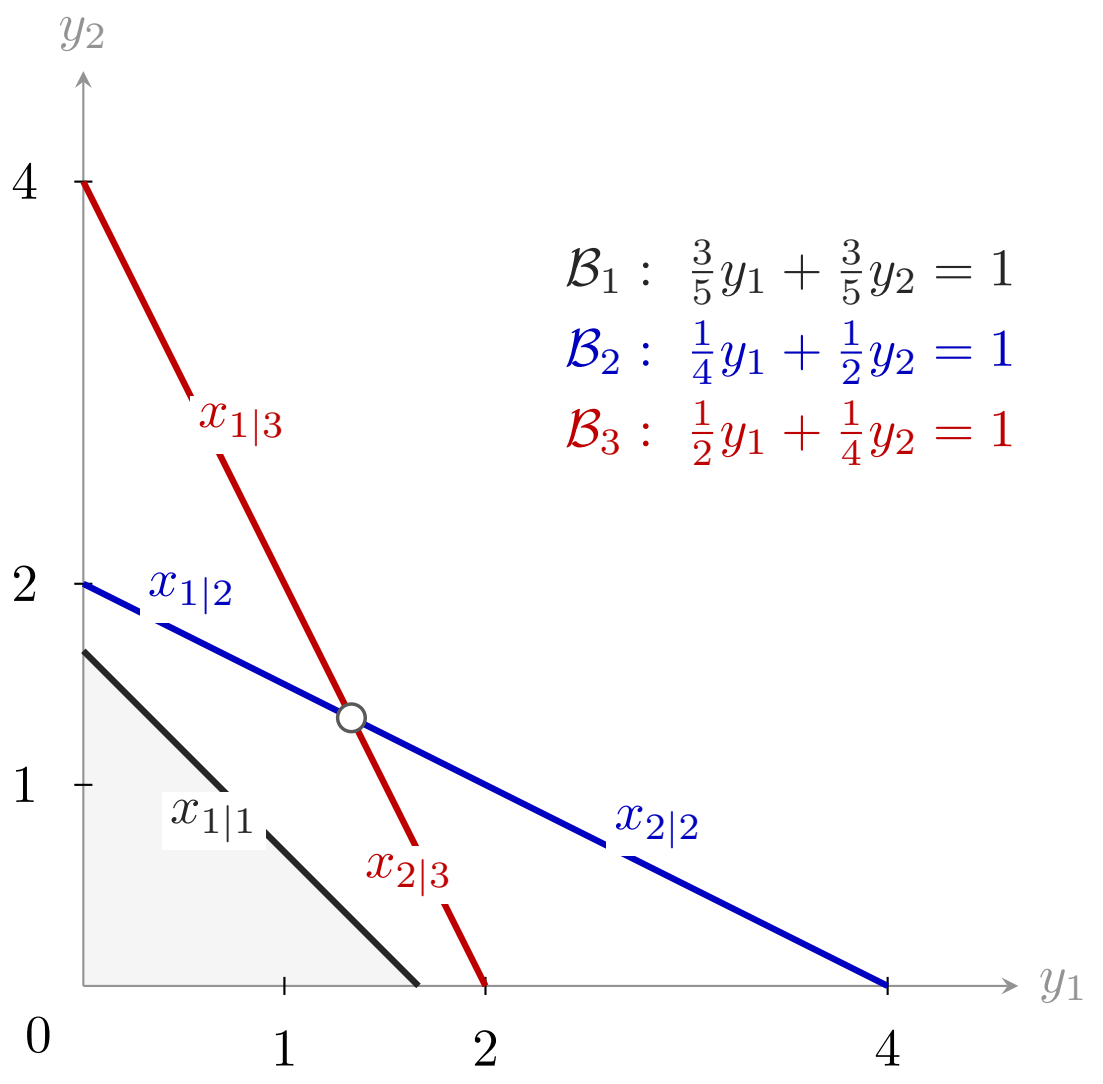
\begin{figure}[htbp]
    \centering
    \begin{tikzpicture}[
        x=1.33cm,y=1.33cm,>=stealth,
        every node/.style={font=\small}
    ]
        % Feasible set enclosed by the inner budget.
        \fill[gray!8]
            (0,0)--({5/3},0)--(0,{5/3})--cycle;

        % Commodity axes.
        \draw[->,gray!85] (0,0)--(4.65,0)
            node[right] {$y_1$};
        \draw[->,gray!85] (0,0)--(0,4.55)
            node[above] {$y_2$};

        \foreach \t in {1,2,4} {
            \draw (\t,0.045)--(\t,-0.045)
                node[below=3pt] {$\t$};
            \draw (0.045,\t)--(-0.045,\t)
                node[left=3pt] {$\t$};
        }
        \node[below left=3pt] at (0,0) {$0$};

        % Budget frontiers.
        \draw[very thick,black!85]
            (0,{5/3})--({5/3},0);
        \draw[very thick,blue!75!black]
            (0,2)--(4,0);
        \draw[very thick,red!75!black]
            (0,4)--(2,0);

        % Excluded intersection patch.
        \filldraw[fill=white,draw=black!65,line width=0.65pt]
            ({4/3},{4/3}) circle[radius=2.6pt];

        % Patch labels, indexed in increasing order of y_1.
        \node[black!85,fill=white,inner sep=1.5pt]
            at (0.65,0.82) {$x_{1|1}$};

        \node[blue!75!black,fill=white,inner sep=1.5pt]
            at (0.54,1.95) {$x_{1|2}$};
        \node[blue!75!black,fill=white,inner sep=1.5pt]
            at (2.86,0.79) {$x_{2|2}$};

        \node[red!75!black,fill=white,inner sep=1.5pt]
            at (0.79,2.79) {$x_{1|3}$};
        \node[red!75!black,fill=white,inner sep=1.5pt]
            at (1.62,0.55) {$x_{2|3}$};

        % Budget equations, with expenditure normalized to one.
        \node[anchor=west,black!85,fill=white,inner sep=1.5pt]
            at (2.35,3.55)
            {$\B_1:\ \frac35 y_1+\frac35 y_2=1$};

        \node[anchor=west,blue!75!black,fill=white,inner sep=1.5pt]
            at (2.35,3.15)
            {$\B_2:\ \frac14 y_1+\frac12 y_2=1$};

        \node[anchor=west,red!75!black,fill=white,inner sep=1.5pt]
            at (2.35,2.75)
            {$\B_3:\ \frac12 y_1+\frac14 y_2=1$};
    \end{tikzpicture}
    \caption{The inner budget $\B_1$ consists of a single patch and
    lies strictly below both outer budgets. Each outer budget has
    two nonintersection patches. The hollow circle marks the
    excluded intersection.}
    \label{fig:three_budget_dominance2}
\end{figure}

\begin{figure}[htbp]
    \centering
    \begin{tikzpicture}[
        >=stealth,
        state/.style={
            draw=black!60,fill=white,rounded corners=3pt,
            minimum width=10mm,minimum height=7mm,
            inner sep=4pt,font=\small
        },
        flow/.style={
            ->,line width=0.85pt,
            shorten >=1.4pt,shorten <=1.4pt
        },
        patchlabel/.style={
            font=\small,fill=white,inner sep=1.4pt
        },
        layerlabel/.style={
            font=\footnotesize,text=black!55
        }
    ]
        % Retain only nodes on feasible source-to-sink paths.
        \node[state] (empty) at (0,0) {$\varnothing$};
        \node[state] (two) at (3.2,1.6) {$\{\B_2\}$};
        \node[state] (three) at (3.2,-1.6) {$\{\B_3\}$};
        \node[state] (twothree) at (6.4,0) {$\{\B_2,\B_3\}$};
        \node[state,minimum width=16mm] (all) at (9.6,0)
            {$\{\B_1,\B_2,\B_3\}$};

        % First remove one of the two outer budgets.
        \draw[flow,blue!75!black] (empty) --
            node[patchlabel,midway,sloped,above=1pt]
            {$x_{2|2}$} (two);
        \draw[flow,red!75!black] (empty) --
            node[patchlabel,midway,sloped,below=1pt]
            {$x_{1|3}$} (three);

        % After budget 2, either patch on budget 3 is admissible.
        \draw[flow,red!75!black] (two) to[bend left=14]
            node[patchlabel,midway,sloped,above=1pt]
            {$x_{1|3}$} (twothree);
        \draw[flow,red!75!black] (two) to[bend right=14]
            node[patchlabel,midway,sloped,below=1pt]
            {$x_{2|3}$} (twothree);

        % After budget 3, either patch on budget 2 is admissible.
        \draw[flow,blue!75!black] (three) to[bend left=14]
            node[patchlabel,midway,sloped,above=1pt]
            {$x_{1|2}$} (twothree);
        \draw[flow,blue!75!black] (three) to[bend right=14]
            node[patchlabel,midway,sloped,below=1pt]
            {$x_{2|2}$} (twothree);

        % Budget 1 must be removed last.
        \draw[flow,black!85] (twothree) --
            node[patchlabel,midway,above=1pt]
            {$x_{1|1}$} (all);

        % Number of budgets already removed.
        \node[layerlabel] at (0,-2.5) {$|S|=0$};
        \node[layerlabel] at (3.2,-2.5) {$|S|=1$};
        \node[layerlabel] at (6.4,-2.5) {$|S|=2$};
        \node[layerlabel] at (9.6,-2.5) {$|S|=3$};
    \end{tikzpicture}
    \caption{The reduced multi-arc powerset digraph for
    Figure~\ref{fig:three_budget_dominance2}. The three unreachable
    nodes have been omitted. Arc colors identify the budget being
    removed, and arc labels identify the chosen patch.}
    \label{fig:dominated-budget-flow}
\end{figure}
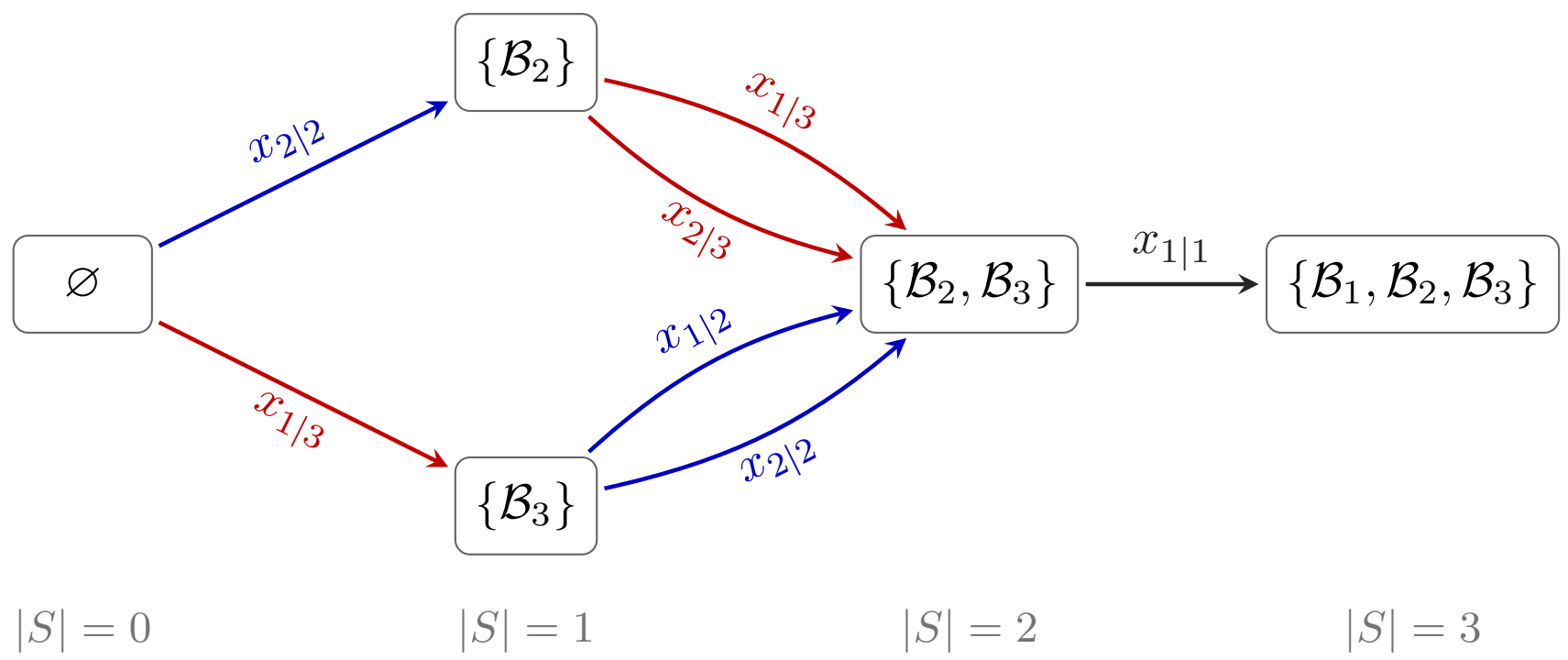

Because each node records the budgets already removed, $\B_1$ can
be removed only after both $\B_2$ and $\B_3$. Consequently, the nodes
$\{\B_1\}$, $\{\B_1,\B_2\}$, and $\{\B_1,\B_3\}$ are unreachable from $\varnothing$
and can be omitted. The reduced network, shown in
Figure~\ref{fig:dominated-budget-flow}, has five nodes and seven arcs.
If $\B_2$ is removed first, its chosen patch must be $x_{2|2}$;
if $\B_3$ is removed first, its chosen patch must be $x_{1|3}$.
Either patch on the second outer budget is then admissible, followed
by the sole patch $x_{1|1}$ on $\B_1$.

We are now ready for formal definitions.

\subsection{Formal Definitions}

Let $\mathcal{B}:=\{\B_1,\ldots,\B_J\}$ denote the collection of budgets.

We first introduce the notion of undominance, following \cite{koida2024dual}.

\begin{definition}[Undominated patch]\label{def:undominated}
Let $x_{i\mid j}$ be a patch on budget $\B_j$, and let
$\widetilde{\mathcal{B}}\subseteq\mathcal{B}\setminus\{\B_j\}$ be a
collection of budgets. We say that $x_{i\mid j}$ is
\emph{undominated} by $\widetilde{\mathcal{B}}$ if it lies strictly
above every budget hyperplane in $\widetilde{\mathcal{B}}$, that is,
\begin{equation}
    p_\ell^\top y>1
    \qquad
    \text{for all }y\in x_{i\mid j}
    \text{ and all }
    \B_\ell\in\widetilde{\mathcal{B}}.
    \label{eq:undominated}
\end{equation}
\end{definition}

Since intersection patches have been excluded, $x_{i\mid j}$ is
undominated by $\widetilde{\mathcal{B}}$ if and only if no patch on
any budget in $\widetilde{\mathcal{B}}$ dominates it in the sense of
the relation $\succ^d$ from Definition~\ref{def:patch_monotonicity}.
By convention, every patch is undominated by the empty collection.

\begin{definition}[Multi-arc powerset digraph]\label{def:ma_digraph}
The initial node set is $2^{\mathcal{B}}$, with source node
$\varnothing$ and terminal node $\mathcal{B}$. For each
$k=0,\ldots,J$, define the layer
\[
    \mathcal{L}_k
    :=
    \{S\subseteq\mathcal{B}:|S|=k\}.
\]
A node $S$ records the budgets already removed, so the remaining
budgets are those in $\mathcal{B}\setminus S$.

The initial layer is $\mathcal{L}_0=\{\varnothing\}$. We construct
the graph iteratively, proceeding through the layers in increasing
order of cardinality. For each $k=0,\ldots,J-1$, each retained node
$S\in\mathcal{L}_k$, and each $\B_j\in\mathcal{B}\setminus S$,
add an arc
\[
    a=(S,S\cup\{\B_j\},i):
    \qquad
    S\xrightarrow{\;x_{i\mid j}\;}S\cup\{\B_j\}
\]
for every $i\in\{1,\ldots,I_j\}$ such that $x_{i\mid j}$ is
undominated by the other remaining budgets
$\mathcal{B}\setminus(S\cup\{\B_j\})$, that is, satisfies
\eqref{eq:undominated} with
$\widetilde{\mathcal{B}}=\mathcal{B}\setminus(S\cup\{\B_j\})$.

After all candidate arcs from layer $\mathcal{L}_k$
to layer $\mathcal{L}_{k+1}$ have been considered, remove every
node in $\mathcal{L}_{k+1}$ with no incoming arcs. 

Let $\mathcal{G}=(\mathcal{V},\mathcal{A})$ denote the resulting
directed multigraph.
\end{definition}

Now we define the unit-flow polytope.
\begin{definition}[Unit-flow polytope]\label{def:ufp}
Let $\mathcal{B}:=\{\B_1,\ldots,\B_J\}$ denote the collection of
budgets, and let $\mathcal{G}=(\mathcal{V},\mathcal{A})$ be its
associated multi-arc powerset digraph. For an arc
$a=(S,S',i)\in\mathcal{A}$, write its flow as $f_a=f_{S,S',i}$.

The \emph{unit-flow polytope} $\mathcal{F}_{\mathcal{G}}$ is the
set of all nonnegative flow vectors $f\in\mathbb{R}_+^{\mathcal{A}}$
satisfying, for every node $S\in\mathcal{V}$,
\begin{equation}
    \sum_{(S',S,i)\in\mathcal{A}} f_{S',S,i}
    -
    \sum_{(S,S',i)\in\mathcal{A}} f_{S,S',i}
    =
    \begin{cases}
        -1, & S=\varnothing,\\
        1,  & S=\mathcal{B},\\
        0,  & \text{otherwise}.
    \end{cases}
    \label{eq:budget-unit-flow-polytope}
\end{equation}
The first sum is the total inflow into $S$. The second sum is
the total outflow from $S$.
A sum over an empty set of arcs is defined to be zero.

We write these constraints compactly as $\mathcal{F}_{\mathcal{G}}=\{f\in\mathbb{R}_+^{\mathcal{A}}:A_{\mathcal{G}}f=b_{\mathcal{G}}\}$, where $A_{\mathcal{G}}$ is the node--arc incidence matrix, and $b_{\mathcal{G}}$ has entry $-1$ at the source, $1$ at the terminal node, and $0$ elsewhere.
\end{definition}

The node--arc incidence matrix $A_{\mathcal{G}}$ is totally unimodular, and $b_{\mathcal{G}}$ is integral. Hence, the unit-flow polytope $\mathcal{F}_{\mathcal{G}}$ is integral. Since $\mathcal{G}$ is acyclic, the vertices of $\mathcal{F}_{\mathcal{G}}$ are precisely the incidence vectors of directed paths from $\varnothing$ to $\mathcal{B}$.

We define the following linear mapping from the flow space to choice probability space.

\begin{definition}[Mapping flows to choice probabilities]\label{def:map}
Let $\mathcal{B}:=\{\B_1,\ldots,\B_J\}$ denote the collection of
budgets, and let $\mathcal{G}=(\mathcal{V},\mathcal{A})$ be its
associated multi-arc powerset digraph, with unit-flow polytope
$\mathcal{F}_{\mathcal{G}}$. For a flow
$f\in\mathcal{F}_{\mathcal{G}}$, define the induced choice-probability
vector $\rho_f$ by
\begin{equation}
    (\rho_f)_{i\mid j}
    :=
    \sum_{\substack{S\subseteq\mathcal{B}\\
        (S,S\cup\{\B_j\},i)\in\mathcal{A}}}
        f_{S,S\cup\{\B_j\},i},
    \label{eq:budget-flow-to-choice}
\end{equation}
for $j=1,\ldots,J$ and $i=1,\ldots,I_j$. Thus, each coordinate
is the total flow on all arcs labeled $x_{i\mid j}$. We denote
the corresponding linear map by $L_{\mathcal{G}}$, so that
$\rho_f=L_{\mathcal{G}}f$.
\end{definition}

In words, for each patch $x_{i|j}$ on budget set $j$, $(\rho_f)_{i|j}$ equals the sum of the flows on all arcs associated with $x_{i|j}$.

Now we are ready to state the network characterization theorem. Recall the definition of $\mathcal{P}^{\textrm{KS}}$ from \eqref{def:KS}.

\begin{theorem}\label{thm:network}
Given a collection of budgets $\mathcal{B}=\{\B_1,\ldots,\B_J\}$,
let $\mathcal{G}=(\mathcal{V},\mathcal{A})$ be its associated
multi-arc powerset digraph from Definition~\ref{def:ma_digraph},
$\mathcal{F}_{\mathcal{G}}$ its associated unit-flow polytope from
Definition~\ref{def:ufp}, and $L_{\mathcal{G}}$ the linear map from
Definition~\ref{def:map}. Then
\begin{equation}
    \mathcal{P}^{\mathrm{KS}}
    =
    \left\{
        L_{\mathcal{G}}f:
        f\in\mathcal{F}_{\mathcal{G}}
    \right\}.
\end{equation}
\end{theorem}
\newpage
\section{Proofs}\label{section:proofs}

We first prove that the dominance relation
 $\succ^d$  defined by Definition~\ref{def:patch_monotonicity} is acyclic. 

\begin{lemma}
     $\succ^d$ is acyclic.
\end{lemma}

\begin{proof}
Suppose, for contradiction, that there exists a cycle of distinct patches
\begin{equation}\label{def:cycle}
    \widetilde{x}_1\succ^d\widetilde{x}_2
    \succ^d\cdots\succ^d\widetilde{x}_m
    \succ^d\widetilde{x}_1.
\end{equation}
Let $\B_b$ denote the budget containing $\widetilde{x}_1$. Since $\widetilde{x}_1\succ^d\widetilde{x}_2$, there exist $y_1\in\widetilde{x}_1$ and $y_2\in\widetilde{x}_2$ such that $y_1\geq y_2$ componentwise. Because the patches are disjoint and $p_b$ is strictly positive,
\begin{equation*}
    p_b^\top y_2<p_b^\top y_1=1.
\end{equation*}
By the definition of patches, $\widetilde{x}_2$ therefore lies entirely strictly below $\B_b$. Furthermore, whenever $x\succ^d x'$ and $x$ lies strictly below $\B_b$, the same argument implies that $x'$ also lies strictly below $\B_b$. Applying this implication successively around the cycle in \eqref{def:cycle} shows that $\widetilde{x}_1$ lies strictly below $\B_b$, contradicting $\widetilde{x}_1\subseteq\B_b$. Thus, $\succ^d$ is acyclic.
\end{proof}
 
\subsection{Proof of Theorem~\ref{thm:projection}}
The third result follows from the first two results and the linearity of coordinate projection. In particular, the first result implies that $\mathrm{Proj}(\mathcal{P}^r)\subseteq\{Av:v\in\Delta^T\}$, and the second result implies that $\{Av:v\in\Delta^T\}\subseteq\mathrm{Proj}(\mathcal{P}^r)$, establishing the desired equality.

We now prove the first result. Let $\pi\in\Pi_r$ be any strict ranking, and write $\succ_\pi$ for the corresponding binary relation. For each $\D_j$, let $x_j^\pi$ denote the selected patch. No patch lying strictly below $\B_j$ can be selected, because every such patch is dominated by at least one patch on $\B_j$ and $\pi$ satisfies patch monotonicity.\footnote{For any strictly affordable bundle $y$, the bundles on $\B_j$ that strictly exceed $y$ in every coordinate form a nonempty relatively open set. This set cannot be covered by the finitely many intersection patches on $\B_j$, each of which has zero $(K-1)$-dimensional measure.}  Thus, $x_j^\pi\subseteq\B_j$ for every $j$. It remains to show that these choices induce a rational type in the sense of KS (i.e., Definition \ref{def:rationaltype}).

Consider the directed revealed-preference graph on the budgets $\{\B_j\}_{j=1}^J$ defined in Definition \ref{def:rp}. Recall that the graph contains an edge $\B_a\to\B_b$ for $a\neq b$ whenever the selected patch $x_b^\pi$
lies strictly below $\B_a$. Suppose, for contradiction, that this graph contains a cycle
\begin{equation}
    \B_{a_1}\to\B_{a_2}\to\cdots\to\B_{a_S}=\B_{a_1}.
\end{equation}
For any edge $\B_a\to\B_b$, we have
$p_a^\top y_b<1=p_a^\top y_a$ for every $y_a\in x_a^\pi$ and $y_b\in x_b^\pi$. Hence, $x_b^\pi\in\D_a$, where $\D_a$ is the expanded choice set associated with $\B_a$. Since $\pi$ selects $x_a^\pi$ from $\D_a$, we have
\begin{equation}
    x_a^\pi\succ_\pi x_b^\pi.
\end{equation}
Applying this argument along the cycle yields
\begin{equation}
    x_{a_1}^\pi\succ_\pi x_{a_2}^\pi
    \succ_\pi\cdots\succ_\pi x_{a_S}^\pi
    =x_{a_1}^\pi,
\end{equation}
contradicting the acyclicity of the strict ranking $\pi$. Therefore, the induced boundary choices satisfy SARP and form a rational type in the sense of KS.

We now prove the second result. We show that every rational type can be extended to a strict preference ranking over patches that satisfies patch monotonicity. Fix a rational type. Let $x_j\subseteq\B_j$ denote the selected patch for $j=1,\ldots,J$. Since the type is rational, its revealed-preference graph on the budgets is acyclic. We can therefore choose a strict ordering consistent with this graph and, after relabeling the budgets, write
\begin{equation}\label{def:budget_ranking}
    \B_1\succ\B_2\succ\cdots\succ\B_J.
\end{equation}
Define a binary relation $\succ^S$ as follows: 
for each $s=1,\ldots,J$, define
\begin{equation*}
    x_s\succ^S x
    \quad\Longleftrightarrow\quad
    x\neq x_s
    \text{ and }
    x\subseteq\B_j
    \text{ for some }j\in\{s,\ldots,J\},
\end{equation*}
where $x_s$ is the selected patch from ${\cal B}_s$.

\begin{comment}
In addition, let $\succ^d$ denote the dominance relation from Definition~\ref{def:patch_monotonicity}: for distinct patches $x,x'\in\mathcal{X}$,
\begin{equation*}
    x\succ^d x'
    \quad\Longleftrightarrow\quad
    \text{there exist }y\in x\text{ and }y'\in x'
    \text{ such that }y\geq y'\text{ componentwise}.
\end{equation*}
\end{comment}

We will show that the union $\succ^S\cup\succ^d$ is acyclic, where $\succ^d$ is defined in Definition~\ref{def:patch_monotonicity}. It can then be extended to a strict preference ranking over $\mathcal{X}$ by applying Theorem~1.4 of \cite{chambers2016revealed} to its reflexive and transitive closure.

\begin{comment}
We next show that $\succ^d$ is acyclic. Suppose, for contradiction, that there exists a cycle of distinct patches
\begin{equation}\label{def:cycle}
    \widetilde{x}_1\succ^d\widetilde{x}_2
    \succ^d\cdots\succ^d\widetilde{x}_m
    \succ^d\widetilde{x}_1.
\end{equation}
Let $\B_b$ denote the budget containing $\widetilde{x}_1$. Since $\widetilde{x}_1\succ^d\widetilde{x}_2$, there exist $y_1\in\widetilde{x}_1$ and $y_2\in\widetilde{x}_2$ such that $y_1\geq y_2$ componentwise. Because the patches are disjoint and $p_b$ is strictly positive,
\begin{equation*}
    p_b^\top y_2<p_b^\top y_1=1.
\end{equation*}
By the definition of patches, $\widetilde{x}_2$ therefore lies entirely strictly below $\B_b$. Furthermore, whenever $x\succ^d x'$ and $x$ lies strictly below $\B_b$, the same argument implies that $x'$ also lies strictly below $\B_b$. Applying this implication successively around the cycle in \eqref{def:cycle} shows that $\widetilde{x}_1$ lies strictly below $\B_b$, contradicting $\widetilde{x}_1\subseteq\B_b$. Thus, $\succ^d$ is acyclic.
\end{comment}

Suppose by way of contradiction that $\succ^S\cup\succ^d$ contains a cycle. Note that the relation $\succ^S$ is acyclic. By Lemma 1, $\succ^d$ is also acyclic. Consequently, the cycle in $\succ^S\cup\succ^d$ must contain edges from both relations. In particular,  the cycle must contain at least one $\succ^S$-edge. Consider a segment between consecutive $\succ^S$-edges of the form
\begin{equation}
    x_{s_i}\succ^S\widetilde{x}_{\tilde{s}_i}
    \succ^d\cdots\succ^d x_{s_k}
    \succ^S\widetilde{x}_{\tilde{s}_k},
\end{equation}
where $\widetilde{x}_{\tilde{s}_i}\subseteq\B_{\tilde{s}_i}$. The two $\succ^S$-edges may be adjacent, with no $\succ^d$-edge in between.

We claim that $s_i<s_k$. By the definition of $\succ^S$ and \eqref{def:budget_ranking}, we have $s_i\leq \tilde{s}_i$. If the $\succ^d$-chain is nonempty, the argument following \eqref{def:cycle} in Lemma 1 implies that $x_{s_k}$ lies strictly below $\B_{\tilde{s}_i}$. Thus, $\tilde{s}_i\neq s_k$ and $\B_{\tilde{s}_i}\to\B_{s_k}$ is a revealed-preference edge. By \eqref{def:budget_ranking}, we therefore have $\tilde{s}_i<s_k$, and hence $s_i\leq \tilde{s}_i<s_k$. If the $\succ^d$-chain is empty, then $\widetilde{x}_{\tilde{s}_i}=x_{s_k}$, so $x_{s_i}\succ^S x_{s_k}$ directly implies $s_i<s_k$.

Applying this argument to the successive $\succ^S$-edges around the cycle yields
\begin{equation*}
    s_1<s_2<\cdots<s_L<s_1,
\end{equation*}
where $x_{s_1},\ldots,x_{s_L}$ are the sources of $\succ^S$ edges. This is a contradiction. Therefore, $\succ^S\cup\succ^d$ is acyclic. 

The reflexive and transitive closure of $\succ^S\cup\succ^d$ is therefore a partial order. By Theorem~1.4 of \cite{chambers2016revealed}, it admits a linear-order extension. Let $\pi$ denote the strict preference ranking given by the strict part of this extension. By construction, $\pi$ satisfies patch monotonicity.

For every $j$, the relation $\succ^S$ places $x_j$ above every other patch on $\B_j$. The argument used in the first part shows that a ranking satisfying patch monotonicity must select a boundary patch from each expanded choice set $\D_j$. Hence, $\pi$ selects $x_j$ from $\D_j$ for every $j$, proving the second result.
\subsection{Proof of Theorem \ref{thm:network}}

As noted after Definition~\ref{def:ufp}, the vertices of
$\mathcal{F}_{\mathcal{G}}$ are the incidence vectors of directed
paths from $\varnothing$ to $\mathcal{B}$. We first show that each
such path $r$ is mapped by $L_{\mathcal{G}}$ to a rational type.

We first show that $L_{\mathcal{G}}r$ is a choice type. Along a path,
each budget $\B_j$ is added exactly once, namely, at the unique
transition
\[
    S\longrightarrow S\cup\{\B_j\}.
\]
The corresponding arc has exactly one patch label $x_{i\mid j}$ for some $i$.
Hence, exactly one patch is selected from each budget $\B_j$.
Therefore, $L_{\mathcal{G}}r$ is a choice type.

We next show that $L_{\mathcal{G}}r$ is a rational type. We verify that $L_{\mathcal{G}}r$ is rational by showing that the induced revealed-preference graph on budgets is acyclic according to Definition \ref{def:rp} and Defition \ref{def:rationaltype}.

To construct such a graph, consider
the patch $x_{i\mid j}$ selected when the path moves from $S$ to
$S\cup\{\B_j\}$. By the definition of the multi-arc powerset
digraph, $x_{i\mid j}$ lies strictly above every budget in
\[
    \mathcal{B}\setminus(S\cup\{\B_j\}).
\]
Hence, if the selected patch $x_{i\mid j}$ lies strictly below some
budget $\B_\ell$, then necessarily $\B_\ell\in S$. 

For each $\B_j$ and $\B_l\in S$ that appears before $\B_j$ along the path $r$,  we add an arc
\[
    \B_l\to\B_j.
\]
Therefore, all arcs point from budgets that appear earlier
on the path to budgets that appear later. The resulting
graph contains every arc induced by the choice type
$L_{\mathcal{G}}r$ under Definition~\ref{def:rp}.

The order in which budgets are added along the path thus provides a
topological ordering of the induced revealed-preference graph on budgets as in Definition \ref{def:rp}.
Consequently, the graph is acyclic, and $L_{\mathcal{G}}r$ is a
rational type by Definition \ref{def:rationaltype}.

We now prove the reverse direction. Let $a$ be any rational type. We show that there exists a path $r$ in $\mathcal{F}_{\mathcal{G}}$ such that $L_{\mathcal{G}}r=a$.

Let $x_{a_j|j}$ denote the patch selected by the rational type $a$ from budget $\B_j$, for $j=1,\ldots,J$. By definition, $a$ induces an acyclic revealed-preference graph on the budgets $\mathcal{B}=\{\B_1,\ldots,\B_J\}$. Hence, this graph admits a topological ordering. Without loss of generality, after relabeling the budgets, write this ordering as
\begin{equation}\label{eqn:ordering}
    \B_1\succ\B_2\succ\cdots\succ\B_J,
\end{equation}
where every revealed-preference edge $\B_j\to\B_k$ satisfies $j<k$.

Let $\mathcal{G}$ be the multi-arc powerset digraph that is induced by $\B=\{\B_1,...,\B_j\}$ according to Definition \ref{def:ma_digraph}.
We now show that there exists a path in $\mathcal{G}$ that follows the ordering in \eqref{eqn:ordering} and selects the patches specified by $a$.

Let $S_1=\varnothing$ and, for $k=1,\ldots,J$, define
\[
    S_k:=\{\B_1,\ldots,\B_{k-1}\}.
\]
We construct the path recursively by moving from $S_k$ to $S_k\cup\{\B_k\}$. Let $x_{a_k|k}$ denote the patch selected by $a$ from budget $\B_k$. We claim that the arc
\[
    S_k \xrightarrow{\;x_{a_k|k}\;} S_k\cup\{\B_k\}
\]
is present in $\mathcal{G}$.

Indeed, such an arc could be absent only if $x_{a_k|k}$ were dominated by the remaining budgets in the sense of Definition~\ref{def:undominated}, that is, if it lay strictly below some remaining budget $\B_\ell$ with $\ell>k$. But then the selected patch from $\B_k$ would lie strictly below $\B_\ell$, implying a revealed-preference edge
\[
    \B_\ell\to\B_k.
\]
This contradicts the topological ordering in \eqref{eqn:ordering}, according to which every revealed-preference edge points from an earlier budget to a later budget. Hence, the required arc is present at every step.

It follows that
\[
    \varnothing
    \to \{\B_1\}
    \to \{\B_1,\B_2\}
    \to \cdots
    \to \mathcal{B}
\]
is a path in $\mathcal{G}$ whose arc labels are precisely
$x_{a_1|1},\ldots,x_{a_J|J}$. Therefore, if $r$ denotes the incidence vector of this path, then
\[
    L_{\mathcal{G}}r=a.
\]

Combining the two directions, the vertices of $\mathcal{F}_{\mathcal{G}}$ map exactly onto the rational types. Since $\mathcal{F}_{\mathcal{G}}$ is the convex hull of its path vertices and linear maps commute with convex hulls, we obtain
\[
    \mathcal{P}^{\mathrm{KS}}
    =
    \left\{
        L_{\mathcal{G}}f:
        f\in\mathcal{F}_{\mathcal{G}}
    \right\}.
\]
\bibliography{references}
\bibliographystyle{plainnat}

\end{document}